%% file: p13pevolv_v06.tex
\documentclass[]{aa}

\usepackage[varg]{txfonts}

\usepackage{lineno}
\usepackage{natbib}
\usepackage{amsmath}
\usepackage{graphicx}
\usepackage{wasysym}
\usepackage{txfonts}
\usepackage{xspace}
\usepackage{url}
\usepackage{textcomp}
\usepackage{multirow}
\usepackage{lipsum}

\newcommand{\ngculx}{NGC\,5907~ULX\xspace}

\newcommand{\pth}{NGC\,7793~P13\xspace}

\newcommand{\swift}{\textsl{Swift}\xspace}

\newcommand{\chandra}{\textsl{Chandra}\xspace}
\newcommand{\xmm}{\textsl{XMM-Newton}\xspace}

\newcommand{\nustar}{\textsl{NuSTAR}\xspace}

\newcommand{\asec}{\ensuremath{''}\xspace}

\newcommand{\snr}{S/N\xspace}

\newcommand{\msun}{\ensuremath{\text{M}_{\odot}}\xspace}

\newcommand{\redchi}{\ensuremath{\chi^{2}_\text{red}}\xspace}

\newcommand{\ergps}{\ensuremath{\text{erg\,s}^{-1}}\xspace}
\newcommand{\heii}{\ensuremath{\rm He\,{\small II}}}

\newcommand{\porb}{\ensuremath{P_\text{orb}}\xspace}

\newcommand{\specpaper}{Walton et al., in prep.} 

\begin{document}

\title{Long-term pulse period evolution of the ultra-luminous X-ray pulsar NGC\,7793 P13}

\author{F.~F\"urst\inst{1}\and D.~J.~Walton\inst{2} \and
M.~Heida\inst{3} \and M.~Bachetti\inst{4} \and C.~Pinto\inst{5}  \and M.~J.~Middleton\inst{6}\and M.~Brightman\inst{7}
\and H.~P.~Earnshaw\inst{7} \and D.~Barret\inst{8}  \and A.~C.~Fabian\inst{2} \and P.~Kretschmar\inst{9} \and K.~Pottschmidt\inst{10,11} \and A.~Ptak\inst{11} \and T.~Roberts\inst{12} \and D.~Stern\inst{13} \and N.~Webb\inst{8} \and J.~Wilms\inst{14}
}

\institute{Quasar Science Resources SL for ESA, European Space Astronomy Centre (ESAC), Science Operations Departement, 28692 Villanueva de la Ca\~nada, Madrid, Spain
\and Institute of Astronomy, Madingley Road, Cambridge CB3 0HA, UK
\and European Southern Observatory, Garching, Germany
\and INAF - Osservatorio Astronomico di Cagliari, via della Scienza 5, I-09047 Selargius, Italy
\and INAF - IASF Palermo, Via U. La Malfa 153, I-90146 Palermo, Italy
\and Department of Physics and Astronomy, University of Southampton, Highfield, Southampton SO17 1BJ, UK
\and Cahill Center for Astronomy and Astrophysics, California Institute of Technology, Pasadena, CA 91125, USA
\and CNRS, IRAP, 9 Av. colonel Roche, BP 44346, F-31028 Toulouse cedex 4, France
\and  European Space Astronomy Centre (ESAC), Science Operations Departement, 28692 Villanueva de la Ca\~nada, Madrid, Spain
\and CRESST, Department of Physics, and Center for Space Science and Technology, UMBC, Baltimore, MD 21250, USA
\and NASA Goddard Space Flight Center, Greenbelt, MD 20771, USA
\and Centre for Extragalactic Astronomy, Department of Physics, Durham University,South Road, Durham DH1 3LE, UK
\and Jet Propulsion Laboratory, California Institute of Technology, Pasadena, CA 91109, USA
\and Dr. Karl-Remeis-Sternwarte and ECAP, Sternwartstr. 7, 96049 Bamberg, Germany
}

\abstract{
Ultra-luminous X-ray pulsars (ULXPs) provide a unique opportunity to study persistent super-Eddington accretion. Here we present the results of a long-term monitoring campaign of ULXP \pth, focusing on the pulse period evolution and the determination of the orbital ephemeris.
Over our four year monitoring campaign with \swift, \xmm, and \nustar, we measured a continuous spin-up with an average value of $\dot P \approx -3.8\times10^{-11}$\,s\,s$^{-1}$. We find that the strength of the spin-up is independent of the observed X-ray flux, indicating that despite a drop in observed flux in 2019, accretion onto the source has continued at largely similar rates. The source entered an apparent off-state in early 2020, which might have resulted in a change in the accretion geometry as no pulsations were found in observations in July and August 2020.

We used the long-term monitoring to update the orbital ephemeris, as well as the periodicities seen in both the observed optical and UV magnitudes and the X-ray fluxes. We find that the optical and UV period is very stable over the years, with $P_\text{UV} = 63.75^{+0.17}_{-0.12}$\,d. The best-fit orbital period determined from our X-ray timing results is $64.86\pm0.19$\,d, which is almost a day longer than previously implied, and the X-ray flux period is $65.21\pm0.15$\,d, which is slightly shorter than previously measured. The physical origin of these different flux periods is currently unknown.

We study the hardness ratio of the \xmm and \nustar data between 2013--2020 to search for indications of spectral changes. We find that the hardness ratios at high energies are very stable and not directly correlated with the observed flux. At lower energies we observe a small hardening with increased flux, which might indicate increased obscuration through outflows at higher luminosities.

Comparing the changes in flux with the observed pulsed fraction, we find that the pulsed fraction is significantly higher at low fluxes. This seems to imply that the accretion geometry already changed  before the source entered the deep off-state.
We discuss possible scenarios to explain this behavior, which is likely driven by a precessing accretion disk.
}


\date{Received XX.XX.XX / Accepted XX.XX.XX}

\maketitle

\section{Introduction}
\label{sec:intro}

The discovery of pulsations from the ultra-luminous X-ray source (ULX), M82 X-2 \citep{bachetti14a}, which led to its identification as an accreting neutron star has opened up a new way of looking at extreme accretion regimes. Such systems, known as ULX pulsars (ULXPs), defy the spherical Eddington limit by orders of magnitude, with the most extreme case being \ngculx with luminosities in excess of $10^{41}$\,\ergps  or about 500 times the Eddington luminosity for a standard neutron star \citep{walton16b, israel17a, ngc5907}. One of the most easily studied ULXPs is \pth \citep[hereafter P13,][]{p13, israel17b}, as it is nearby \citep[$d=3.40\pm0.17$\,Mpc, ][]{zgirski17a}, is isolated from other sources in its host galaxy, and exhibits (almost) persistent pulsations. P13 has a pulse period of around 415\,ms and typical luminosities of around $5\times10^{39}$--$10^{40}\,\ergps$, clearly placing it  in the ULX regime, which is typically defined as $L_x > 10^{39}\,\ergps$.  \citet{p13} measured a spin-up of $\dot P \approx -3.5\times10^{-11}$\,s\,s$^{-1}$ which they used to infer a dipole magnetic field of around $1.5\times10^{12}$\,G based on the accretion model of \citet{ghosh79a}.

The source  was initially discovered in the X-rays by \citet{read99a}. \citet{motch11a}  identified the companion and mass donor as a B9Ia super-giant.  Later, \citet{motch14a} found an optical and UV photometric period of $\approx$64\,d, which is also present in the radial velocity of the \heii\ emission. While the origin of the \heii\ emission line is debated \citep{fabrika15a}, \citet{motch14a} interpreted the clearly detected period as  the orbital period  of the system and find a dynamical mass constraint of  $<$15\,\msun for the compact object, ruling out an intermediate-mass black hole and providing evidence for super-Eddington accretion in the system.
% (which was a very small upper limit at the time, as the neutron star nature of the compact object was not known).

\citet[hereafter F18]{p13orb} used accurate X-ray period measurements obtained with \xmm and \nustar to constrain all parameters of the orbital ephemeris of P13 and found an orbital period of $63.9\pm0.5$\,d (statistical uncertainties only), thus confirming the results from \citet{motch14a}. 
They could also constrain the eccentricity $\epsilon$ to be very small ($\epsilon \leq 0.14$). This almost circular orbit  is in slight contradiction with the larger eccentricity implied from the optical light curve, which was necessary to explain the narrow optical maximum under the assumption that the compact object is a black hole \citep{motch14a}. Updated calculations based on more recent optical data and assuming a neutron star accretor might resolve those differences.

The  X-ray flux also shows large variations in a number of different timescales. One important periodic variability is found around  $65.05\pm0.1$\,d \citep{hu17a}, based on long-term \swift/XRT monitoring data.  This period modulates the flux by a factor of 3--4 during the bright state of P13. Using a longer baseline, F18 updated the results of \citet{hu17a}  and found  an X-ray period of $66.8\pm0.4$\,d. The difference between the optical and UV as well as the X-ray values might be due resonances in the accretion disk or caused by a warped and precessing accretion disk \citep[F18]{hu17a}. This super-orbital period could also explain the variation in the arrival times of maximum light in the optical  \citep{motch14a, hu17a}. 

P13 also shows strong long-term X-ray flux variations, for example it exhibits X-ray off-states, where its flux drops below the detection limit of \swift/XRT.  On the other hand, it has been in a long, bright X-ray flux state since at least 2016 and likely even since 2013, though we lack dense flux  monitoring before 2016. Typical luminosities during this time were around $10^{40}$\,\ergps.  Between January and March 2019 it entered a low state, with the flux dropping drastically over the next few months until it was briefly no longer detectable in individual XRT snapshots \citep{soria20aAtel, hu20aAtel}, before recovering to a low, but significantly detected flux.  It is currently unknown if these long-term flux variations are periodic or random.

In July 2020, we obtained a \chandra observation of P13 to obtain a measurement of the low state flux \citep{walton20aATel}. We find a luminosity of $\left(4.1\pm0.5\right)\times10^{38}$\,\ergps  (in the 0.3--10\,keV band) with a spectrum consistent with the one obtained from the low-state in 2011 and 2012, but with a flux at least an order of magnitude higher\footnote{To put this in context, this low state flux is still at the upper end of fluxes typically observed from X-ray binaries in the Milky Way.}. Based on contemporaneous \swift/XRT monitoring, it seems that the source had already left the deepest off-state during the \chandra observation. Notably this implies that the current off-state was much shorter than the one in 2011 and 2012, which lasted $\gtrsim$2\,years. 

The \chandra data do not provide sufficient time-resolution to measure the pulse period of P13, and are therefore not analyzed here. 
However, the increased flux encouraged us to ask for further monitoring with \xmm and \nustar in July and August 2020. While the flux was higher, neither of those observations yielded detectable pulsations. A full analysis of those data will be presented in a forthcoming publication (\specpaper). \swift monitoring through October 2020 shows that P13 continues to be active at a low level.

Here we report on continued \nustar and \xmm monitoring, using new data taken in 2018, 2019, and 2020, and following the flux, spectral, and pulse period evolution into the renewed off-state. In Sect.~\ref{sec:data} we describe the observations analyzed here and the data reduction methods. In Sect.~\ref{sec:analysis}, we discus the data analysis, including the UV and X-ray flux period (Sect.~\ref{susec:uvxrayper}), the pulse period and its evolution (Sect.~\ref{susec:pulses}), the evolution of the hardness ratios (Sect.~\ref{susec:spectra}) and the behavior of the pulsed fraction as a function of time and spectral parameters (Sect.~\ref{susec:pf}). We discuss the results in Sect.~\ref{sec:summ} and conclude the paper with a  summary  and outlook in Sect.~\ref{sec:conlc}.
Uncertainties are given at the 90\% confidence level, unless otherwise noted.

\section{Observations and data reduction}
\label{sec:data}

\input{perevol3_2020_evt_3-10keV.tex}

\subsection{\swift}

\begin{figure*}[]
\begin{center}
\includegraphics[width=0.95\textwidth]{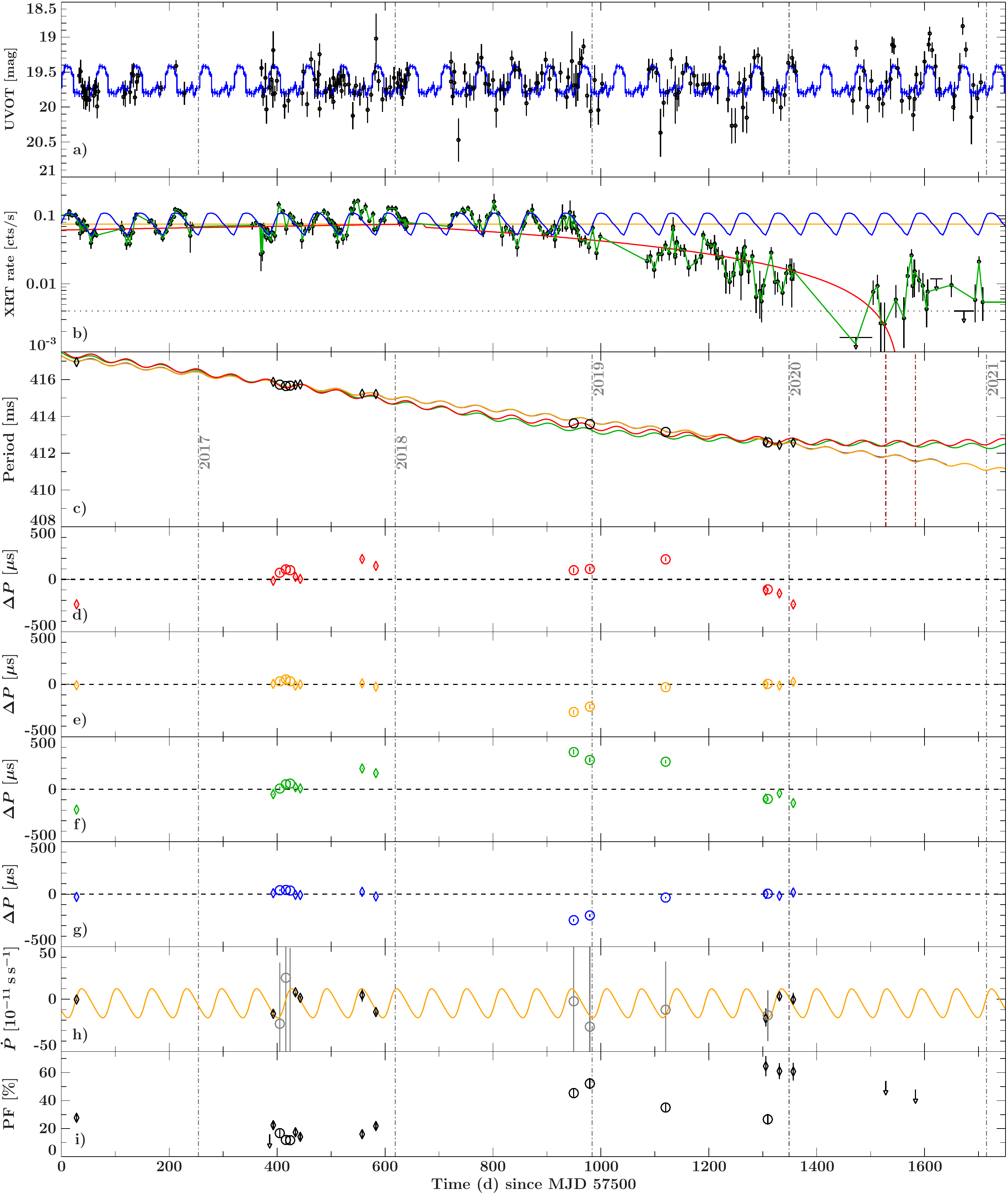}
\caption{ \textit{a)} \swift/UVOT light curve in the $U$-band.  \textit{b)} \swift/XRT lightcurve. The colored lines show the four different assumed flux evolutions, which are used as an input for the pulse period evolution:  constant flux (orange),  linear brightening and dimming trend (red),  measured XRT lightcurve (green), and extrapolated X-ray super-orbital period profile (blue). The dotted line represents the estimated count rate at the Eddington limit. \textit{c)} Pulse period evolution as measured by \xmm (circles) and \nustar (diamonds). Superimposed are the best-fit models for the four different input lightcurves in the same colors as in panel \textit{b)}. The brown dotted-dashed lines indicate the times of observations when no pulsations where seen. 
  \textit{d)} Residuals with respect to the linear brightening and dimming input, \textit{e)} residuals with respect to a constant input, \textit{f)} residuals with respect to the original lightcurve as input, and \textit{g)} residuals with respect to the X-ray profile input.  \textit{h)} Measured (black  and gray) and predicted (orange) pulse period derivative $\dot P$. The model is based on the constant flux model. \textit{i)} Pulsed fraction in the 3--10\,keV energy band. Upper limits are denoted by downward pointing arrows. For details see text.}
\label{fig:lc_xrt_uvot}
\end{center}
\end{figure*}

Since April 2016, P13 has been monitored by the \textsl{Neil Gehrels Swift Observatory} \citep[\swift;][]{swiftref} in each of its visibility windows, with a typical cadence of around one week or less and exposure times of 1\,ks per snapshot. The visibility constraints result in five observation epochs, each lasting for around nine months (see Fig.~\ref{fig:lc_xrt_uvot}).
Results from previous \swift monitoring data are discussed by F18.

In addition to the data presented by F18, we extracted 131 XRT \citep{swiftxrtref} observations taken between 2018-04-14 (ObsID 00093149031) and 2020-12-31 (ObsID 00031791109) with the standard \swift/XRT processing pipeline \citep{evans09a}, thereby extending the data presented by F18 by over three years.
The data are binned such that there is a single 0.3--10\,keV flux
measurement  from each observation. Selected observations during the low-state at the end of 2020 were combined manually to yield more stringent upper limits.  

We also extracted UVOT \citep{swiftuvotref} data from all 131 new observations, following the same method as detailed by F18.  In particular, we used a circular source region with a  5\asec radius centered on $\alpha$ = 23h $57' 50.9''$, $\delta$ = $-32^\circ37'26.6''$ and a 15\asec circular background region. The data were processed with the corresponding software tasks as distributed by HEASOFT v6.24 and we used \texttt{uvotsource} to extract the source magnitudes.

Figure~\ref{fig:lc_xrt_uvot} shows the long-term light curve of these observations obtained with the UVOT (panel a) and XRT (panel b) instruments. These light curves are further discussed in Sect.~\ref{susec:uvxrayper}.

\subsection{NuSTAR}

Data from \nustar \citep{harrison13a}  were reduced using the standard pipeline, \texttt{nupipeline} and \texttt{nuproducts}, provided
with the \nustar Data Analysis Software (v1.8.0), using standard filtering and \nustar CALDB v20191219.
We extracted source events from both focal plane models (FPMA and FPMB) in circular regions with a radius of 35\asec and background events from circular regions with a radius of 120\asec on the same detector as the source.
We chose the source region size based on optimizing the signal-to-noise ratio (\snr) above 10\,keV. Larger regions will add disproportionately more background photons than source photons, reducing the high energy \snr.
All time information was transferred to the solar barycenter using the DE-200 solar system ephemeris \citep{solarDE200}.
To search for pulsations we combined the source filtered event files for FPMA and B to improve the statistics.

\subsection{XMM-Newton}

Data from \xmm \citep{xmmref}  were reduced with the \xmm Science Analysis System (SAS) v18.0.0,
following the standard prescription\footnote{http://xmm.esac.esa.int/}.
We only use data from EPIC-pn  \citep{pnref} in this work, as it provides the necessary fast time resolution to search for pulsations.
The  data were taken in full frame mode and raw data  files were cleaned and calibrated using \texttt{epchain} and transferred to the solar barycenter using the SAS task ``barycen'' based on the DE-200 solar system ephemeris \citep{solarDE200}.

We extracted source events for all epochs from circular regions with a radius of 40\asec, following the same method as described in F18. Background spectra were extracted from a source free circular region with a radius of $\sim$100\asec, located on the same CCD as P13. We carefully checked all observations for background flaring, but found that it was only problematic for epoch 2017A, which prevents us from measuring the pulse period in that observation (as discussed in F18). See Table~\ref{tab:perevol} for a complete observation log.

\section{Analysis}
\label{sec:analysis}

\subsection{UV and X-ray periods}
\label{susec:uvxrayper}
\begin{figure}
\begin{center}
{\includegraphics[width=0.95\columnwidth]{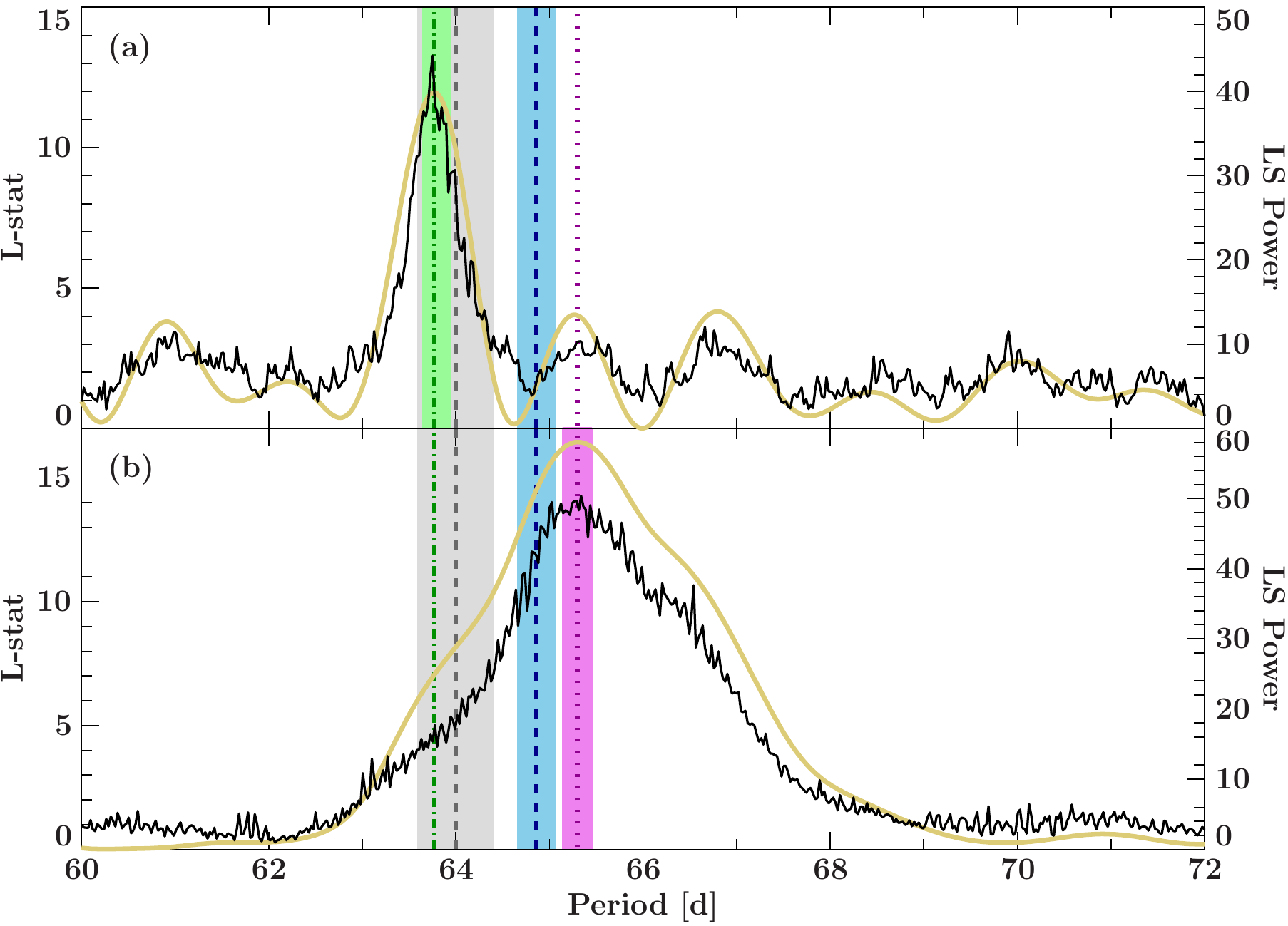}}
\caption{Results from epoch folding (black left $y$-axis) and Lomb-Scargle periodogram (yellow, right $y$-axis) for the UV light curve (top) and X-ray light curve (bottom). The strongest UV period is marked by the green dotted-dashed line, the strongest X-ray period by the purple dotted line. The new orbital period is indicated by the blue dashed line, the old estimate of the orbital period is shown by the gray dashed line. The shading behind each period indicates their respective uncertainties (using the updated uncertainties for the old orbital period in gray).}
\label{fig:epfold}
\end{center}
\end{figure}

Given the much longer timeline of available \swift monitoring data, we updated the long-term periods presented by F18. We used the same approach as presented in F18, that is to say we performed epoch folding \citep{leahy83a} and calculated a Lomb-Scargle periodogram  \citep{scargle82a} for both the \swift/UVOT and the \swift/XRT light curve (Fig.~\ref{fig:epfold}). For epoch folding, we used the L-statistics proposed by \citet{davies90a} for increased sensitivity.

Due to the high variability in flux (see Fig.~\ref{fig:lc_xrt_uvot}\textit{b}), we neededto normalize the XRT data. F18 used a linear brightening trend and removed it from the data. As such a trend is obviously no longer a good fit, we instead opted to renormalize each epoch to its respective mean count-rate. This approach is the same as used by \citet{hu17a}. 
No renormalization was done for the UVOT data given their overall stability.
Uncertainties (at the 90\% level)  were determined by simulating 5000 light curves, sampled with the same cadence as the real light curve, with each point drawn randomly from a Poisson distribution based on an interpolation of the respective  folded profile.

We find an optical period of $P_\text{UV} = 63.75^{+0.17}_{-0.12}$\,d (Fig.~\ref{fig:epfold}, top), in very good agreement to previous results \citep[F18]{motch14a}. In the X-rays, we find a period of $P_\text{X}=  65.31\pm0.15$\,d (Fig.~\ref{fig:epfold}, bottom), significantly shorter than the $66.8 \pm 0.4$\,d value reported by F18. However, the value we measure here is close to the one presented by \citet{hu17a}: $P_\text{X,Hu}= 65.05\pm0.1$.  Even with this reduction, the X-ray period is very significantly different from the optical period.
We checked that the method of removing the underlying variability does not influence the measured value, that is to say we obtain the same results for renormalizing each epoch, subtracting a trend, or not changing the data at all. However, the statistical detection of the X-ray period is significantly improved when using the renormalization for each epoch.

We note that the UV period is much more pronounced during the X-ray low-state in 2020. A continuation of the UV period even during X-ray low-states was already discussed by \citet{motch14a}, who attributed it to the fact that a large precessing accretion disk shields the X-rays from us, but not towards the companion star. The UV variability would then be caused by the X-ray heated side of the companion periodically turning towards us. However, this does not necessarily explain why the UV variability is suppressed during the X-ray high state. A dilution of the UV period due to stronger contribution from the accretion disk to the UV flux seems unlikely, as the average $U$-band magnitude of the system did not change during the X-ray high state.

While the peak in the X-ray periodogram appears much broader than the peak in the UV periodogram, we do not find any evidence that the X-ray period is quasi-periodic in nature. By splitting the data into smaller parts, we find no indication that the X-ray period is changing in value, however it is more pronounced during the X-ray high state.

\subsection{Pulsations}
\label{susec:pulses}

\subsubsection{Pulsation search}

We searched for pulsations in all new observations (2018A--2020C), using the same accelerated epoch folding search as used in F18. In particular, we searched  for pulsations over a grid in the plane defined by the  pulse period, $P$, and its first time derivative, $\dot P$, with the data binned into 12 phase bins. To limit the search ranges, we used the secular spin-up and orbital ephemeris found by F18 as an estimator for the expected pulse period during each observation. We then performed a search around that estimated period in a 100$\times$100 grid with $\Delta P = \pm 0.3$\,ms and $\Delta \dot P = \pm 2\times10^{-9}$\,s\,s$^{-1}$. Due to their longer duration, the \nustar observations provide more constraining measurements (in particular for $\dot P$). We therefore performed a second search for the \nustar data only, where we zoomed in on the peak found in the previous calculation and searching a 120$\times$120 grid with $\Delta P = \pm 0.1$\,ms and $\Delta \dot P = \pm 6\times10^{-11}$\,s\,s$^{-1}$. We found highly significant pulsations with a significance $> 99.5\%$ in all epochs but 2020B and 2020C.  
Those data were taken during the recovery from the off-state and we discuss them in more  detail below.
The significance is based on a $\chi^2$-statistic corrected for the number of trials corresponding to the bins in the $P$-$\dot P$  grid. 

To estimate an upper limit on the pulsed fraction in the three epochs where no pulsations are detected (2017A, 2020B, and 2020C), we simulate event files based on an input light curve with the same average count rate as the real data, but with added sinusoidal pulsations, resulting in a pulsed fraction $\text{PF}_\text{sim}$. The simulations are based on the method used in the \texttt{Stingray} package \citep{huppenkothen19a}.
For each observation, we simulate event lists with pulsed fractions, $\text{PF}_\text{sim}$, $0.1\le \text{PF}_\text{sim}\le 0.9 $ in steps of 0.01. For each value of $\text{PF}_\text{sim}$ we generate 100 statistically independent event lists and search each list for pulsations using the epoch folding technique.
We do not include $\dot P$ in these simulations as it does not influence the obtained upper limit. We define the upper limit on the pulsed fraction as the value where at least 90\% of all simulated event lists provide a detection of the pulse with at least 99.5\% significance. The results are listed in Table~\ref{tab:perevol}. We estimated the upper limits assuming a constant pulsed fraction over the whole energy band of the respective instrument.

Uncertainties on $P$ and $\dot P$ were determined from the extent of the full-width half-maximum (FWHM) contour in the 2D $\chi^2$ landscape of the epoch folding results. That is, we define the uncertainties in both parameters as the range where the 2D $\chi^2$ peak has dropped to half of its peak value.

The measured values for $P$ and $\dot P$ for each observation are given in Table~\ref{tab:perevol} and plotted in Fig.~\ref{fig:lc_xrt_uvot}\textit{c}. As can be seen the source continued its secular spin-up with roughly $\dot P \approx -3.8\times10^{-11}$\,s\,s$^{-1}$ (corresponding to $\dot \nu \approx 2.21\times10^{-10}$\,Hz\,s$^{-1}$ in frequency space) before pulsations were no longer detected in mid 2020. The measured spin-up in each observation is found to vary on the order of a few $\pm10^{-10}$\,s\,s$^{-1}$ around the average value, as the orbital Doppler effect dominates there.

\subsubsection{Pulse period evolution and orbital ephemeris}

We describe the pulse period evolution with a combination of secular spin-up and orbital motion. We apply the same model as described in F18, which allows us to fit for the orbital parameters (orbital period \porb, eccentricity $\epsilon$, projected semi-major axis $a \sin i$, argument of periastron $\omega$, and time of periastron $\tau$) and requires as input a term related to the accretion of angular momentum. Our first order  assumption was that the observed X-ray flux should be an adequate tracer of the accreted angular momentum, following  standard accretion theory \citep{ghosh79a, ghosh79b}. In this description the pulse period change is expected to be proportional to $PL^{6/7}$, where $L$ is the (bolometric) luminosity.  As we do not know the exact coupling constant or conversion between observed X-ray count-rates and luminosity, we subsume these conversion in one factor, the spin-up parameter $b$ \citep[for details see][and F18]{marcu15b, Bissinger2016}.

One of the main issues with using the measured X-ray flux as tracer of accreted angular momentum is that our observation history of the X-ray has gaps that can last weeks to months (see Fig.~\ref{fig:lc_xrt_uvot}\textit{b}). These are mainly due to gaps in visibility of the source for \swift, and therefore occur roughly once a year. Furthermore the observed X-ray flux may be modulated by intrinsic absorption or changing of the beaming factor of the emitted X-rays, while the actually accreted mass and angular momentum has not changed.

F18 circumvented the problem of the missing data by replacing the measured X-ray flux with two simple models: a linear brightening trend and a variable profile based on folding the \swift/XRT data on the 66.9\,d super-orbital X-ray period.  

The new data show that a linear brightening trend is no longer a realistic description of the long-term light curve, given the large drop in observed flux in 2019. We therefore modify the trend  with a break at around MJD 58300, after which a linear dimming trend is applied (red model in Fig.~\ref{fig:lc_xrt_uvot}). This approach allows us to build on the solution for the orbit and pulse period evolution found by F18, but also captures the overall shape of the long-term light curve.

However, this model fails to explain the full data-set, leaving large residuals in the X-ray timing data (Fig.~\ref{fig:lc_xrt_uvot}\textit{d}). The best fit implies an orbital period of around 65\,d (formal uncertainty calculation is not feasible here given the overall bad quality of the fit). Separately, the data before and after January 2018 can be fitted well, however, the best-fit solutions seem to be incompatible with each other, with  $\porb^\text{old} = 64\pm0.4$\,d and $\porb^\text{new} =61.0^{+0.7}_{-0.6}$\,d. Compared to F18, we find slightly larger uncertainties on the orbital period in the data before 2018. Upon closer investigation we found that the uncertainties reported in F18 are underestimated due to a bug in the minimization routine, which has since been fixed.

We also note that the model proposed by Ghosh \& Lamb, and in particular the assumptions about how the magnetic field connects to the accretion disk, are likely not applicable in the case of ULXPs. For example, the extreme accretion rates in ULXPs  will lead to the formation of geometrically thick accretion disks which were not discussed by \citet{ghosh79b}. Interestingly, \citet{p13} found  that the Ghosh \& Lamb theory can explain the spin-up of P13 with a magnetic field of around $1.5\times10^{12}$\,G; however, this only works for the high luminosities observed in 2013--2016. With the lower luminosities observed in 2019, the model predicts much lower maximal spin-up rates, independent of the magnetic field.

A better description of the overall pulse period evolution is obtained when assuming a constant X-ray flux as input (orange model in Fig.~\ref{fig:lc_xrt_uvot}), that is to say a constant secular spin-up only modulated by the orbital period. This approach implies that the observed X-ray luminosity is not tracing the accretion of angular momentum. 
This model leaves small residuals around the densely sampled epoch in 2017 ($t\approx400$\,d in Fig.~\ref{fig:lc_xrt_uvot}), however, it provides a much better match to the most recent data during the low flux state of the source. We find an orbital period of around 64.9\,d.

We find the same general behavior when comparing a model using the directly measured XRT light curve (green in Fig.~\ref{fig:lc_xrt_uvot}) as input vs an input based on the super-orbital X-ray profile with a constant average flux (blue in Fig.~\ref{fig:lc_xrt_uvot}). The large reduction in flux in 2019 in the measured XRT light curve leads directly to an over-prediction of the observed pulse period, while the constant average flux of the profile input provides a much better description of the long-term behavior.

We base our updated orbital calculation on the assumption of a constant spin-up, as it seems to describe the observed observations of the pulse periods best. However, there are still significant outliers in late 2018 ($t\approx1000$\,d in Fig.~\ref{fig:lc_xrt_uvot}) which cannot be explained with this simple model. They are likely caused by brief periods of enhanced accretion, however, they occur at the end of a densely sampled interval, making it unlikely that we missed large X-ray flares that would result in a significant amount of additionally accreted matter and angular momentum. On the other hand, because the X-ray flux is not a good tracer for the amount of accreted angular momentum, it is possible that a spin-up due to enhanced accretion occurred without leaving a measurable trace in the X-ray lightcurve.
For calculating the updated ephemeris we therefore first ignore those data points, and discuss the impact of different scenarios to describe them below. 

The overall fit of this model is still not very good in terms of $\chi^2$, with $\chi^2 =  64.7$ for 7 degrees of freedom (based on 7 orbital parameters and 14 data points). To allow realistic error calculation, which requires a $\chi^2\approx1$, we add 0.005\% of systematic uncertainties on all measurements of the pulse periods (which implies a factor 2--5 increase over the statistical uncertainties and is likely related to timing noise), resulting in a reduced $\chi^2$, $\redchi$, of 1.06 for the same number of degrees of freedom (dof). Including the ``outlier'' data around MJD\,58500 results in a best-fit with only a $\chi^2=88.3$ for 8 dof even with those systematic uncertainties.

\begin{figure*}
\begin{center}
{\includegraphics[width=0.95\textwidth]{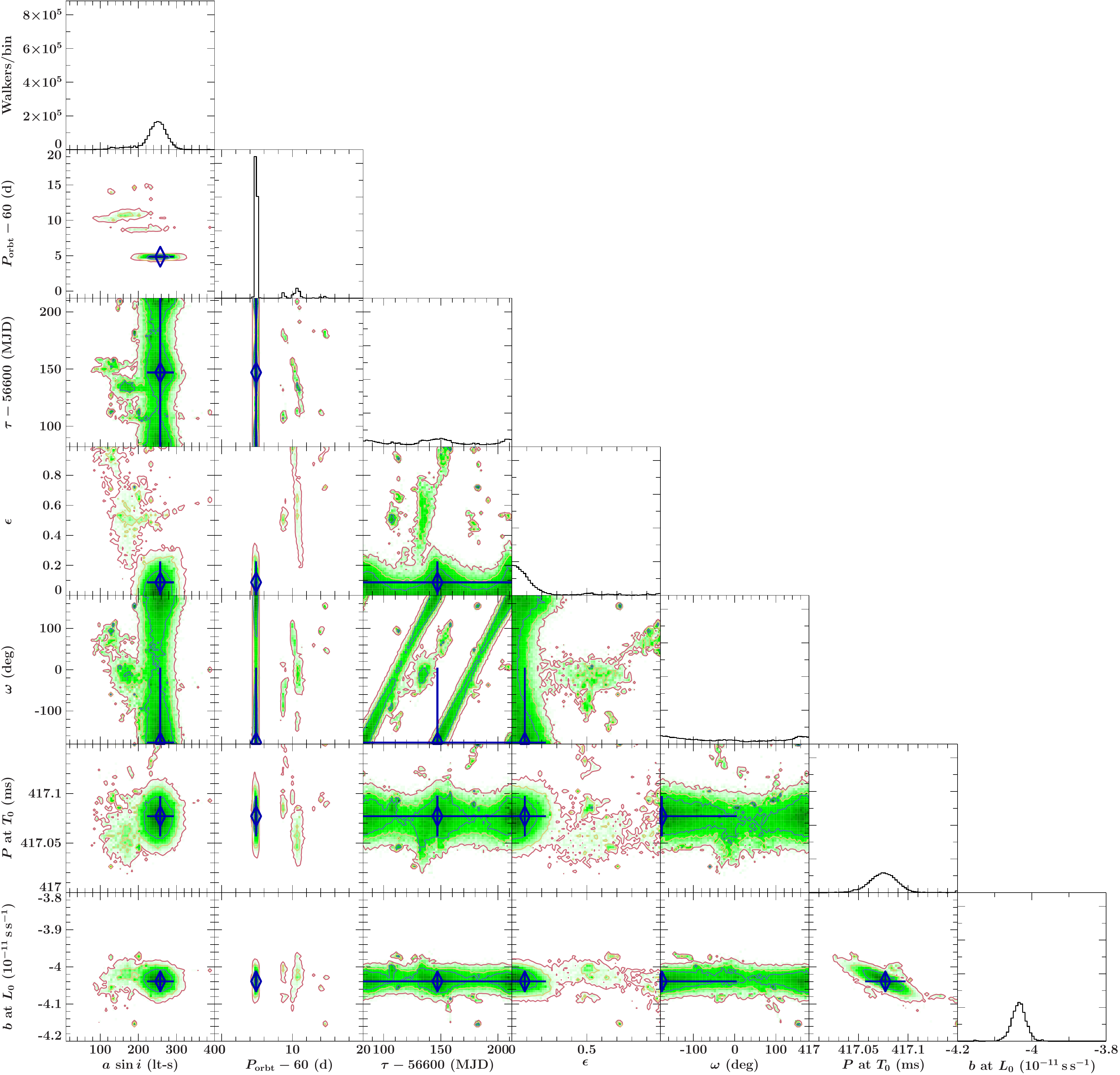}}
\caption{Orbital parameters distributions based on the results from the MCMC run using a constant spin-up model and ignoring the two measurements around MJD\,58500. The blue diamonds and error-bars indicate the results from a standard $\chi^2$ optimization. The contours show the 68\%, 90\%, and 99\% confidence intervals.}
\label{fig:mcmc}
\end{center}
\end{figure*}

Given the complexity of the fit and the low number of degrees of freedom, we also run MCMC simulations to estimate the posterior distribution of each parameter. We use an implementation of the ``emcee'' sampler \citep{foreman12a} in ISIS, which is based on the method proposed by \citet{goodman10a}. We use 210 walkers (30 walker per free parameter) and evolve them for 3000 steps. Before calculating the distributions of walkers we use a 20\% burn-in period. The results are shown in Fig.~\ref{fig:mcmc}, together with the best-fit values and uncertainties from the standard $\chi^2$-optimizer.

We find that spin-up strength and initial pulse period (at MJD\,57530.0) are very well constrained. We find a best-fit orbital period of $64.87^{+0.52}_{-0.27}$\,d, which is almost a day longer than the orbital period presented by F18 and implied by \citet{motch14a}. The orbital period shows a weak secondary maximum of around 61\,d, which also corresponds to a slightly smaller projected semi-major axis and a much larger eccentricity, which seems unphysical and in particular does not describe the densely sampled 2017 data well. We therefore ignore this minimum and report the 1D uncertainties  for the orbital period only based on the main peak at 64.87\,d. 

This longer orbital solution compared to the one presented by F18 is necessary to explain the behavior of the pulse period in late 2019 and early 2020 ($t\approx1350$\,d in Fig.~\ref{fig:lc_xrt_uvot}). These new data strongly constrain the orbital phase, highlighting how important a dense sampling is for constraining the orbital period. With an orbital period of 63.9\,d as found by F18, we find that the phase is almost half a period off. While it is possible that the orbital period changes in this system due to loss of angular momentum \citep[see, e.g.,][]{bachetti20a}, the required change would be orders of magnitude larger than expected. We find, however that the older F18 estimate and the updated constraints on the orbital period presented here are  are still marginally consistent within their $\sim$99\% uncertainties. 

The argument of periastron, $\omega$, is basically unconstrained, which is a result of the vanishing eccentricity, $\epsilon$, which is consistent with 0 (similar to the results by F18). Overall, the results from the MCMC run agree well with the values obtained by $\chi^2$ fitting. We present the 1D uncertainties from the parameter distributions in Table~\ref{tab:orbfit}.

As mentioned above, this new ephemeris is obtained when ignoring two measurements at the end of 2018. Clearly, the source underwent some stronger spin-up over the course of 2018 than predicted by our model. To test the influence of those data points on our ephemeris, we split the data in two parts, one before January 2018 and one after. We then require that both parts have the same orbital solution, but allow for different spin-up and $P(0)$ values between them. With this, we basically allow a rapid spin-up event at some point during 2018 and possible lower spin-up trend from December 2018 to 2020. 
We find that the orbital parameters using this model are fully compatible with the values when ignoring the 2018 data. In particular, we find $P=65.05\pm0.25$\,d, which is consistent with the orbital period in the previous model and also significantly longer than the UV/optical period.

Regarding the spin-up, we find $\dot P_1 = \left(-3.93\pm0.11\right)\times10^{-11}$\,s\,s$^{-1}$ for the first part, and $\dot P_2=\left(-3.37\pm0.13\right)\times10^{-11}$\,s\,s$^{-1}$ for the second part. Both are lower then the best-fit solution presented in Table~\ref{tab:orbfit} as this model has an implicit jump of $\Delta P$ of around $-0.7$\,ms sometime in 2018. More observations in the future are required to constrain if $\dot P$ did indeed change in 2018.

\input{orbfits_table_2020.tex}

\subsection{Spectral evolution}
\label{susec:spectra}

In many accreting sources, large changes in flux go together with significant changes in the spectral shape, including X-ray pulsars in the Milky Way \citep[see, e.g.,][]{reig18a}. As many of the spectra show a rather low signal-to-noise ratio (\snr), we restrict ourselves to studying hardness ratios as a proxy for spectral change. A more detailed spectral analysis will be presented in a forthcoming publication (\specpaper).

\begin{figure}
\begin{center}
{\includegraphics[width=0.95\columnwidth]{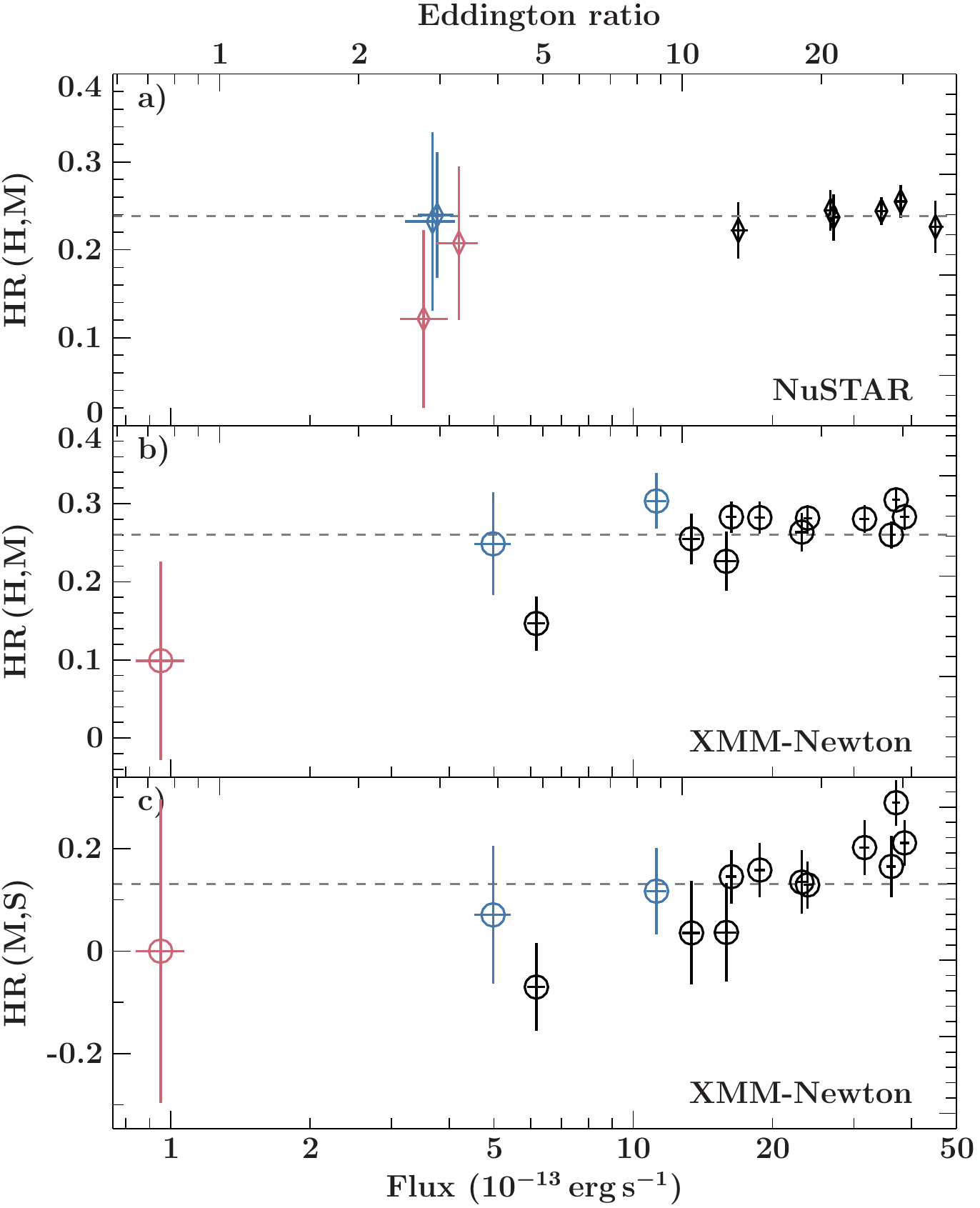}}
\caption{Hardness ratio as a function of flux for all \xmm and \nustar data. \textit{a)} HR between the 5.0--10.0\,keV and 3.0--5.0\,keV energy band for \nustar. \textit{b)} HR between the same bands as in (a) but for \xmm. \textit{c)} HR between the 3.0--5.0\,keV and 0.5--1.5\,keV band  for \xmm. Data taken in 2020 are shown in red, data from 2019 in blue, and all previous data in black. The dashed lines show the respective average HR for each instrument for all data before 2019. The top $y$-axis gives the flux as a fraction of the Eddington limit for a 1.4\,\msun neutron star at a distance of 3.4\,Mpc.}
\label{fig:hr2flux}
\end{center}
\end{figure}

We define three energy bands, soft (S) between 0.5--1.5\,keV, medium (M) between 3.0--5.0\,keV, and hard (H) between 5.0--10.0\,keV.  These bands were chosen by eye as they highlight the observed features most clearly, but the exact change of the energy bands does not  influence the overall behavior. The soft band is only available for the \xmm data. We measured the flux in each of these bands based on the spectrum in each observation.
We define the hardness ratio (HR) as
\begin{equation}
\text{HR}(X,Y) = \frac {X - Y } {X + Y}  \quad,
\end{equation}
where $X$ and $Y$ are the fluxes in the harder and softer energy band, respectively. 
We plot the hardness ratio as a function of flux in Fig.~\ref{fig:hr2flux}. As can be seen, there is very little variation in the high energy spectrum, with HR(H,M) almost constant over the whole flux range. At lower energies, a slight hardening with increased flux is visible. This could either be due to an increase in absorption or an intrinsic change in the spectral shape (a more in depth analysis of the spectral evolution will be presented in Walton et al., in prep.). We speculate that at higher luminosities, stronger outflows are launched from the super-Eddington accretion disk, which  contribute to a larger absorption column. 

During the lowest observed flux (\xmm in epoch 2020C) the source was just around the Eddington limit for a 1.4\,\msun neutron star at a distance of 3.4\,Mpc \citep{zgirski17a}. Here the luminosity is estimated based on the 3--10\,keV flux, which is roughly a factor of 2 lower than the bolometric (0.5--100\,keV) accretion luminosity of P13.
At these low flux levels, the top two panels of Fig.~\ref{fig:hr2flux} suggest a slight softening of the spectrum at higher energies.  However, a significant change cannot be claimed given the large measurement uncertainties

\subsection{Pulsed fraction}
\label{susec:pf}

We calculated the pulsed fraction (PF) in all observations in the 3--10\,keV energy band, based on the pulse profile with 12 equally spaced phase bins. We estimate the PF as 
\begin{equation}
\text{PF} = \frac{ \max(\text{PP})-\min(\text{PP}) }{\max(\text{PP})+\min(\text{PP})}
\end{equation}
where PP is the pulse profile. The uncertainty of the PF is based on Gaussian error propagation, which is justified as each bin of the pulse profiles contains at least 25 counts.

We find that during the latest \nustar observations (epochs 2019B, 2019C, and 2020A) the pulsed fraction was significantly higher than in other observations, reaching up to  60\% in the 3--10\,keV band. We show the pulsed fraction as function of time in Fig.~\ref{fig:lc_xrt_uvot}. As shown by F18, the pulsed fraction is typically strongly energy dependent, with higher energies showing higher pulsed fractions. The energy dependence is most significant at low energies (covered \xmm) and levels off at higher energies (covered by \nustar).  The energy dependence is consistent in most observations, with the exception of the observations in epoch 2019B, which have the highest pulsed fraction overall. In this epoch the pulsed fraction is already very high at low energies and does not show a significant energy dependence.

\begin{figure}
\begin{center}
{\includegraphics[width=0.95\columnwidth]{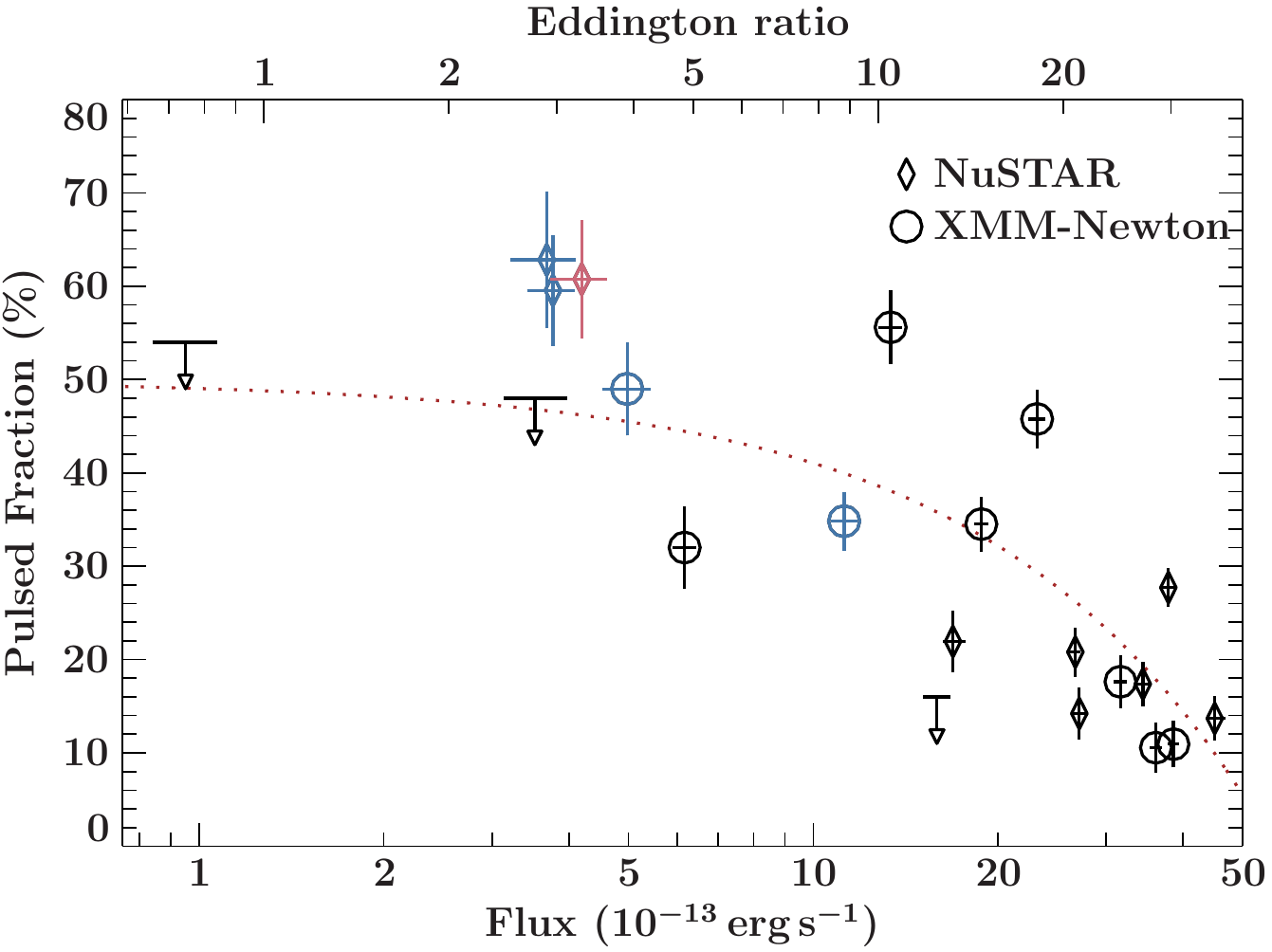}}
\caption{Pulsed fraction in the 3--10\,keV band as a function of flux in the same energy band. \nustar data are shown as diamonds, \xmm data are shown as circles. Data from 2019 are shown in blue, data from 2020 are shown in red. The brown dotted line shows a simple linear regression to the data. The top $y$-axis gives the flux as a fraction of the Eddington ratio for a 1.4\,\msun neutron star at a distance of 3.4\,Mpc.}
\label{fig:flx2pf}
\end{center}
\end{figure}

The pulsed fraction shows an anticorrelation with flux, as shown in  Fig.~\ref{fig:flx2pf}, with a Pearson's correlation coefficient of $-0.83\pm0.07$. We estimated the uncertainty on the correlation coefficient via a bootstrapping resampling method using 10,000 iterations. Using the Student's $t$-test, we find that the anticorrelation is significant at the $>99.9\%$ level.

\begin{figure}
\begin{center}
{\includegraphics[width=0.95\columnwidth]{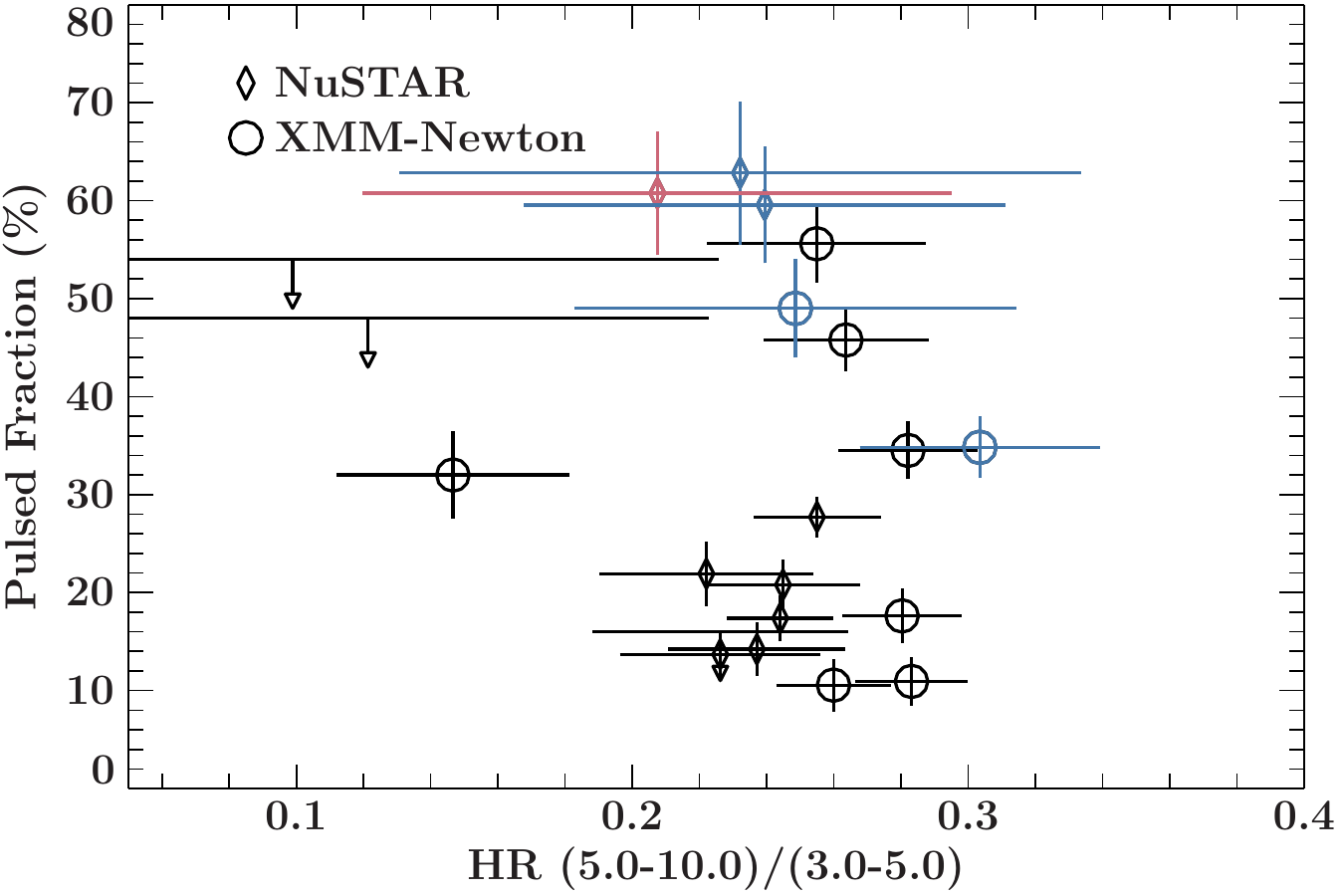}}
\caption{Pulsed fraction in the 3--10\,keV band as a function of the hardness ratio between the 5.0--10.0\,keV and the 3.0--5.0\,keV energy band. \nustar data are shown as diamonds, \xmm data are shown as circles. Data from 2019 are shown in blue, data from 2020 in red.}
\label{fig:hr2pf}
\end{center}
\end{figure}

On the other hand, we do not find a strong correlation overall between the spectral shape, as measured by the hardness and the pulsed fraction (Fig.~\ref{fig:hr2pf}). When taking only the 2019 and 2020 data into account, a correlation can be implied, though it is not statistically significant.

\section{Discussion}
\label{sec:summ}

We have presented an analysis of the X-ray pulsations seen from \pth between 2016--2020. During this period, the source was mostly in an active state, and showed a constant long-term spin-up. However, in 2019 the observed flux faded significantly, dropping below the detection threshold for our \swift/XRT monitoring, before recovering to a more stable (but still low) flux level. Despite this flux evolution, our X-ray timing results imply that the long-term spin-up continued at a similar rate to that seen in the high-flux state. 

In addition to tracking the timing data, we have also explored whether the strength of the pulsed signal evolves with both flux and spectral hardness of P13. The pulsations appear to be strongest at low fluxes, but we find little evidence for any dependence on spectral shape. These results also allow us to make a preliminary assessment of the spectral evolution of P13 across this period. Although during the high-flux period we find little evidence for large spectral variations, we do see some interesting hysteresis associated with the more recent flux evolution. 

 Using a constant spin-up approximation we have updated the orbital ephemeris of \pth and find values inconsistent with the ones presented by F18. In particular, we find an orbital period of $64.86\pm0.19$\,d based on extensive MCMC simulations. This period is  larger than the best-fit orbital period presented by F18, and also longer than the periodicity seen in the UV. On the other hand, it is very close to the revised X-ray period, which we find is $65.31 \pm 0.15$\,d. It is currently unclear how to interpret these different periods in a physical context, and it is particularly puzzling how the optical flux seems to vary on time-scales faster than the orbital period.

 It is possible that our estimate of the orbital period has larger systematic uncertainties than implied. As discussed, not all measured periods fit the curve well; in particular, the two measurements around MJD\,58450 cannot be reconciled with any simple spin-up model. Hence it is possible that there is timing noise present or that there are unobserved spin-up or -down episodes that we cannot model. Adding ad-hoc flares in the gaps of the XRT monitoring, it is possible to find a model describing these points well,  under the assumption that this modified X-ray flux is related to the spin-up value.  However, we still find that an orbital period of around 65\,d is required to describe all the data and this ad-hoc flux evolution is not based on any observational evidence.

The decoupling between the observed spin-up and the X-ray flux could indicate that strong obscuration occurs during the drop in flux in 2019, while the intrinsic accretion is continuing unabated (or shows flares that result in short-term spin changes). This behavior and scenario is similar to the one proposed for NGC\,300~ULX-1, another highly variable ULXP \citep{vasilopoulos20a}. However, if absorption and obscuration is the reason for the diminishing X-ray flux, we would expect to see a significant hardening of the observed X-ray spectrum, which is not the case (Fig.~\ref{fig:hr2flux}).  In fact, we find rather the opposite behavior:  the source is getting softer at lower fluxes.

We find a clear anticorrelation between the pulsed fraction and the source flux in the 0.5--20\,keV energy band, with pulsed fractions as high as 60\% during the low states in 2019 and early 2020.  This anticorrelation is clearly present even outside the lowest fluxes.
It seems to indicate that at lower fluxes the accretion column, which is responsible for the pulsed flux, dominates. According to \citet{walton18a}, the pulsed flux of the accretion column can be described by a power law with an exponential cut-off at high energies. 
Typically this component dominates at higher energies, with non-pulsed emission likely associated with the accretion disk also seen at lower energies. These results may suggest a lower relative contribution in the observed bands from the disk. This will be explored in more detail in future work.
 
 On the other hand, the pulsed fraction does not show a significant correlation with spectral hardness (Fig.~\ref{fig:hr2pf}). We would expect a strong correlation if indeed the pulsed fraction increases because the hard accretion column starts to dominate.  
 Instead, the change pulsed fraction might be related to a changed scattering time within the cone of the accretion disk and wind. This cone confines the emitted X-rays to its opening angle, causing so-called ``beaming''. Before photons emitted from the neutron star can escape this cone, they might undergo a number of scatterings, causing a significant delay in their arrival time. For a large enough cone this might lead to a smeared out pulse profile with a lower pulsed fraction. We expect large accretion disk cones and larger beaming fractions at higher luminosities, providing a possible way to explain the correlation between pulsed fraction and flux without a significant change in the spectral shape.

\section{Conclusion \& outlook}
\label{sec:conlc}

\pth continues to surprise us with new behavior.  It is one of only two ULXPs for which the companion star is identified (the other one being NGC\,300~ULX-1, \citealt{heida19a}, while NGC 1313~X-2 has a known optical counterpart, but the origin of the optical emission is not yet identified, \citealt{sathyaprakash19a, grise08a}), and the only ULXP for which the full ephemeris can be determined. However, the details of this ephemeris are still unclear. With the most recent data the best-fit orbital period is 64.9\,d almost a day longer than the optical and UV period, and about 0.5\,d shorter than the X-ray period. Further observations of the pulse period evolution will allow us to obtain a better understanding if this difference in periods is real or due to a systematic effect in our measurement. 

We have also found a correlation between the flux and the pulsed fraction and have shown that the pulsed fraction can change significantly without a measurable change in spectral shape. Forthcoming detailed spectral modeling (\specpaper) will allow us to investigate this behavior in more detail and probe different scenarios of obscuration by neutral or highly ionized material.
In addition, continued measurement of the pulsed fraction at different flux levels will allow us to fill in the parameter space and investigate if clear changes in accretion geometry occur at certain fluxes. 

A major step forward in our understanding of ULXPs would be provided by updated models no how the torque of the accreted material is transferred onto the neutron star and how the magnetic field couples with the accretion disk in the case of geometrically thick, super-Eddington accretion disks.

\begin{acknowledgements}

We would like the thank the referee for the very useful comments that helped to improve the manuscript.
%MJM
%We thank M.~K\"uhnel  and M.~Nowak for the useful discussions.
DJW and MJM  acknowledge support from STFC Ernest Rutherford fellowships.
%, ACF
%acknowledges support from ERC Advanced Grant 340442, and DB acknowledges
%financial support from the French Space Agency (CNES).
This research has made
use of data obtained with \nustar, a project led by Caltech, funded by NASA and
managed by NASA/JPL, and has utilized the \texttt{nustardas} software package, jointly
developed by the ASDC (Italy) and Caltech (USA). This research has also made
use of data obtained with \xmm, an ESA science mission with instruments and 
contributions directly funded by ESA Member States. This work made
use of data supplied by the UK Swift Science Data Centre at
the University of Leicester, and also made use of the XRT
Data Analysis Software (XRTDAS) developed under the responsibility
of the ASI Science Data Center (ASDC), Italy.
This research has made use of a collection of ISIS functions (ISISscripts) provided by ECAP/Remeis observatory and MIT (\url{http://www.sternwarte.uni-erlangen.de/isis/})
The material is based upon work supported by NASA under award number 80GSFC17M0002.

\end{acknowledgements}

%\textit{Facilities:} \facility{NuSTAR}, \facility{XMM}, \facility{Swift}

%
%\bibliographystyle{jwaabib}
%\bibliography{mnemonic,aa_abbrv,fx_statistics,fx_catalogs,accretion,xrays,radproc,sat,nustar,spin,bh,hmxb,herx1,mypapers,ulx,ns,fx_vela,wd,stars}

\end{document}

%% file: perevol3_2020_evt_3-10keV.tex
\begin{table*}
\caption{Observation log together with their fluxes,  pulse periods, and pulse period derivatives. Data above the horizontal line were already presented in \citet{p13orb}; new data are below the line. For clarity, we also list the epoch labels for the archival data given in \citet{p13} and \citet{walton18a}. The last column gives the pulsed fraction (PF) in the 3--10\,keV band.}
\label{tab:perevol}
\centering
\begin{tabular}{llllllll}
\hline\hline
Mission  & Epoch & ObsID & Date [MJD] & Flux\tablefootmark{a}  & P [ms] $-$415.0\,ms & $\dot P$ [$10^{-10}$\,s\,s$^{-1}$]   & PF [\%]   \\\hline
\xmm & X2/2013 & 0693760401  & 56621.21 & $6.20\pm0.27$ &$4.712\pm0.008$ & $0.2^{+3.4}_{-2.8}$ & $32\pm5$\\
\xmm & X3/2014 & 0748390901  & 57002.00 & $18.9\pm0.6$ &$3.390^{+0.007}_{-0.008}$ & $-0.5^{+3.0}_{-2.5}$ & $34.5\pm3.0$\\
\xmm & XN1/2016 & 0781800101  & 57528.58 & $37.2\pm0.8$ &$1.951^{+0.008}_{-0.007}$ & $0.1^{+2.6}_{-2.9}$ & $23.7\pm2.1$\\
\nustar & XN1/2016 & 80201010002  & 57528.18 & $37.4^{+1.2}_{-0.8}$ &$1.9515^{+0.0016}_{-0.0019}$ & $-0.04^{+0.19}_{-0.17}$ & $27.7\pm2.1$\\
\xmm & 2017A & 0804670201  & 57886.17 & $16.0\pm0.9$ & --- & --- & $<16$  \\
\xmm & 2017B & 0804670301  & 57893.66 & $16.4\pm0.5$ &$0.864^{+0.009}_{-0.006}$ & $-1.1^{+1.6}_{-3.2}$ & $17.5\pm2.9$\\
\nustar & 2017B & 30302005002  & 57892.71 & $17.0^{+0.9}_{-0.7}$ &$0.8755\pm0.0020$ & $-1.39^{+0.27}_{-0.22}$ & $22\pm4$\\
\xmm & 2017C & 0804670401  & 57904.90 & $31.8\pm0.8$ &$0.724\pm0.010$ & $-2\pm6$ & $17.6\pm2.9$\\
\xmm & 2017D & 0804670501  & 57916.10 & $38.8\pm0.9$ &$0.649^{+0.016}_{-0.025}$ & $2^{+13}_{-9}$ & $10.9\pm2.5$\\
\xmm & 2017E & 0804670601  & 57924.11 & $36.3\pm0.9$ &$0.669^{+0.008}_{-0.019}$ & $-6^{+12}_{-5}$ & $10.6\pm2.7$\\
\nustar & 2017F & 30302015002  & 57933.93 & $34.3^{+1.1}_{-1.3}$ &$0.7050^{+0.0024}_{-0.0017}$ & $0.65^{+0.22}_{-0.36}$ & $17.4\pm2.4$\\
\hline
\nustar & 2017G & 30302015004  & 57942.93 & $27.1^{+0.9}_{-1.2}$ &$0.7409^{+0.0023}_{-0.0011}$ & $0.13^{+0.16}_{-0.25}$ & $14.2\pm2.8$\\
\nustar & 2017H & 90301326002  & 58057.58 & $45.1^{+2.0}_{-1.1}$ &$0.2284^{+0.0035}_{-0.0030}$ & $0.4\pm0.7$ & $13.7\pm2.5$\\
\xmm & 2017I & 0804670701  & 58083.00 & $24.0\pm0.6$ &$0.214^{+0.007}_{-0.006}$ & $-1.0^{+2.4}_{-2.6}$ & $25.8\pm2.5$\\
\nustar & 2017I & 30302005004  & 58082.95 & $26.6\pm1.0$ &$0.2153^{+0.0018}_{-0.0024}$ & $-1.23^{+0.31}_{-0.24}$ & $20.8\pm2.7$\\
\xmm & 2018A & 0823410301  & 58449.79 & $23.3\pm0.8$ &$-1.378^{+0.013}_{-0.014}$ & $-0^{+11}_{-10}$ & $46\pm4$\\
\xmm & 2018B & 0823410401  & 58479.60 & $13.4\pm0.6$ &$-1.420^{+0.011}_{-0.014}$ & $-3^{+11}_{-10}$ & $56\pm4$\\
\xmm & 2019A & 0840990101  & 58619.99 & $11.3\pm0.6$ &$-1.843\pm0.009$ & $-1\pm5$ & $35\pm4$\\
\nustar & 2019B & 50401003002  & 58805.68 & $3.7\pm0.5$ &$-2.365\pm0.004$ & $-1.8^{+1.0}_{-0.8}$ & $63\pm8$\\
\xmm & 2019B & 0853981001  & 58809.36 & $5.0\pm0.5$ &$-2.414\pm0.007$ & $-1.5^{+2.4}_{-2.5}$ & $49\pm6$\\
\nustar & 2019C & 30502019002  & 58830.98 & $3.7\pm0.4$ &$-2.5499^{+0.0025}_{-0.0018}$ & $0.27^{+0.22}_{-0.29}$ & $60\pm6$\\
\nustar & 2020A & 30502019004  & 58856.54 & $4.1\pm0.5$ &$-2.4213^{+0.0025}_{-0.0035}$ & $-0.0^{+0.7}_{-0.5}$ & $61\pm7$\\
\xmm & 2020B & 0861600101  & 59027.98 & $1.24^{+0.11}_{-0.09}$ & --- & --- & $<54$ \\
\nustar & 2020C & 90601327002  & 59083.22 & $3.6^{+0.5}_{-0.4}$ & --- & --- & $<48$ \\\hline
\end{tabular}
\tablefoot{
\tablefoottext{a}{Flux in $10^{-13}$\,erg\,s$^{-1}$\,cm$^{-2}$ in the 3--10\,keV energy band.}
}
\end{table*}

%% file: orbfits_table_2020.tex
\begin{table*}
\caption{Best-fit orbital parameters as presented by F18 (left columns) and in this work (right columns), using either a $\chi^2$ minimization method or the MCMC estimator. These results are based on the assumption of a constant spin-up, independent of the observed X-ray flux. The first column uses the same data and model as presented by F18 but using a corrected fit algorithm, resulting in significantly larger uncertainties. All uncertainties are reported at the 90\% level.} \label{tab:orbfit}
\centering
\begin{tabular}{r|ll||ll}
\hline\hline
&   \multicolumn{2}{c||}{F\"urst et al., 2018} &\multicolumn{2}{c}{This Work} \\
 Parameter & $\chi^2$ min.  &  MCMC  & $\chi^2$ min.    & MCMC  \\\hline
 $ P_\text{spin}~[\text{ms}]$ & $417.068^{+0.017}_{-0.020}$ & $417.032^{+0.013}_{-0.019}$ & $417.077\pm0.021$ & $417.075^{+0.013}_{-0.025}$ \\
 $ \dot P~[10^{-11}\,\text{s\,s}^{-1}]$ & $-4.03\pm{0.06}$ & $3.65^{+0.05}_{-0.06}$ & $-4.039^{+0.029}_{-0.030}$ & $-4.050^{+0.026}_{-0.025}$ \\
 $ a\sin i~[\text{lt-s}]$ & $262\pm20$ & $209^{+18}_{-19}$ & $260\pm40 $ & $250^{+40}_{-90}$ \\
 $ P_\text{orb}~[\text{d}]$ & $63.76^{+0.34}_{-0.30}$ & $63.9^{+0.6}_{-0.5}$ & $64.86\pm0.19$ & $64.89^{+0.18}_{-0.22}$ \\
 $ \tau$ ~[MJD] & $56767^{+23}_{-43}$ & $56669^{+26}_{-21}$ & $56750\pm70$ & $56615^{+60}_{-70}$ \\
 $ e$ & $\le0.12$ & $\le 0.14$ & $\le 0.24$ & $\le0.7$ \\
 $ \omega~[\text{deg}]$ & $0^{+180}_{-180}$ & $-10^{+100}_{-110}$ & $-177^{+182}_{-4}$ & $153^{+16}_{-324}$ \\
$\chi^2/\text{d.o.f.}$   & 9.90/2 & & 7.41/7\\\hline
\end{tabular}

\end{table*}